\documentstyle[graphicx,aps,twocolumn]{revtex}

\begin{document}
\draft

 \twocolumn[\hsize\textwidth\columnwidth\hsize  
 \csname @twocolumnfalse\endcsname              

\title{Observation of  resonance trapping in an open microwave
cavity}
\author{E. Persson$^{1,2}$, I. Rotter$^3$, H.-J. St\"ockmann$^1$, and
M. Barth$^1$}
\address{
$^1$Fachbereich
Physik, Philipps-Universit\"at Marburg, Renthof 5, D-35032
Marburg, Germany\\
$^2$Institut f\"ur Theoretische Physik, Technische
Universit\"at Wien, A-1040 Wien, Austria\\
$^3$Max-Planck-Institut f\"ur Physik komplexer
Systeme, D-01187 Dresden, Germany}
\date{July 10, 2000}
\maketitle

\begin{abstract}
The coupling of a quantum mechanical system to open decay channels
has been theoretically studied in numerous works, mainly in the
context of nuclear physics but also in atomic, molecular and
mesoscopic physics. Theory predicts that with increasing coupling
strength to the channels the resonance widths of all states should
first increase but finally decrease again for most of the states.
In this letter, the first direct experimental verification of this
effect, known as resonance trapping, is presented. In the
experiment a microwave Sinai cavity with an attached waveguide
with variable slit width was used.
\end{abstract}

\pacs{PACS numbers: 05.45.-a, 03.65.Nk, 84.40.Az, 85.30.Vw}

 ]  

Since more than ten years, interference phenomena in open quantum
systems have been studied theoretically in the framework of
different models. Common to all these studies is the appearance of
different time scales as soon as the resonance states start to
overlap see \cite{moldauer} and the recent papers \cite{trap}
with references therein).
Some of the states align with the decay channels and
become short-lived while the remaining ones decouple to a great
deal from the continuum and become long-lived (trapped). Due to
this phenomenon, the number of relevant states will, in the
short-time scale, be reduced while the system as a whole  becomes
dynamically stabilized. The phenomenologically introduced doorway
states in nuclear physics provide an example for  the alignment of
the short-lived states with the channels \cite{doorway}.
Calculations for microwave resonators showed that the trapped
resonance states can be identified in the time-delay function and
that short-lived collective modes are formed at large openings of
the resonator \cite{pepirose}. Resonance narrowing is inherent
also in the Fano formalism \cite{meier}. Similar effects have been
found in the linewidths in a semiconductor microcavity with
variable strength of normal-mode coupling \cite{semcav}. In spite
of the many theoretical studies, the effect of resonance trapping
has not yet been verified unambigously in an experiment.
A theoretical study of neutron resonances in nuclei
as a function of the interaction  of a doorway state with
narrow resonances \cite{brentano}
allowed only to draw the conclusion that resonance trapping is not
in contradiction with experimental data.
For a clear experimental demonstration of the trapping effect, the
coupling strength to the decay channels should be tunable, which was not
possible in all above mentioned experiments.

The mechanism of resonance trapping can be illustrated best on the
basis of a schematical model. In an open quantum system the
resonance states are allowed to decay, i.\,e. their energies are
complex, ${\cal E}_R = E_R - \frac{i}{2} \Gamma_R$. The Hamilton
operator is non-hermitian,
\begin{equation}
{\cal H} = {\cal H}_0 - i \alpha V
V^\dagger \; .
\label{eq:heff}
\end{equation}
Here ${\cal H}_0$ describes the
$N$ discrete states of the closed quantum system coupled to $K$
decay channels by the $N\times K$ matrix $V$. ${\cal H}_0$ and
$ V V^\dagger$ are hermitian and $\alpha$ is a real parameter for
the total coupling strength between the closed system and the
channels. The complex eigenvalues of ${\cal H}$ give the energy
positions $E_R$ and  widths $\Gamma_R$ of the resonance states.
Studies on the basis of this model were
performed by different groups with different assumptions on the properties of
${\cal H}_0$ and $V$, see e.g. \cite{chem,sok,rot}. The Hamiltonian ${\cal
H}_0$ can be given by, e.g., a Gaussian orthogonal ensemble or a Poissonian
othogonal ensemble and the elements of $V$ may  follow from
different assumptions on their statistical distribution. In all cases,
the results show clearly the formation of different time scales due to
the anti-hermitian part of ${\cal H}$. At a certain critical value
of the coupling parameter $\alpha$, the widths of $N-K$ states
start to decrease with increasing $\alpha$ and approach zero with
$\alpha \to \infty$ while the widths of $K$ states increase.

The widths of the  states of realistic systems  show a more
complicated behaviour than described in the statistical model. The widths of
the long-lived states of molecules saturate with increasing coupling strength
to the continuum, but do not approach zero \cite{sat1}.
A saturation of the widths of the trapped
states occurs, however, also in the calculations with the
schematical model when it is improved by considering different coupling
strengths for the different decay
channels \cite{sat2}. This improvement, as well as the
introduction of a complex coupling parameter instead of the real
$\alpha$, are justified since they follow from formally rewriting
the Schr\"odinger equation in the function space of both discrete
{\it and} scattering states \cite{nucl}. Similar results are
obtained for nuclei \cite{nucl} and for open microwave cavities
\cite{pepirose}. On the basis of this result it is possible to
clarify another problem discussed for resonance states in
molecules at high level density \cite{sat1}: The fundamental
quantum mechanical relation between the average width of the
states and the average lifetime is violated when all resonance
states are considered
in the averaging procedure. The mechanism of resonance trapping
makes, however, the averaging procedure meaningful only for either
the long-lived or the short-lived states. Performing the average
over the trapped states only, the relation between
the average width and the average lifetime is recovered.

While $\Gamma_R \to 0$ with $\alpha \to \infty$ is a necessary
condition for trapped states in the schematical model with the
Hamiltonian (\ref{eq:heff}), the calculations for realistic
systems show that the widths of the trapped states may decrease,
stop to increase and even slowly increase with increasing coupling
strength to the continuum. Only the ratio between the widths of
the trapped  states and the widths of the short-lived states
decreases with increasing coupling strength according to these
studies. This behaviour may lead  to such an interesting effect as
population trapping resulting from the coupling of two states of
an atom by means of a strong laser field \cite{marost}. The
interplay between the direct interaction of two states and the
real and imaginary parts of their interaction via the continuum
causes unexpected and sometimes contra-intuitive effects. In any
case, the local processes between individual resonances are
of central importance for the properties of the system.

\begin{figure}
\noindent
\includegraphics[width=2.8cm,angle=-90]{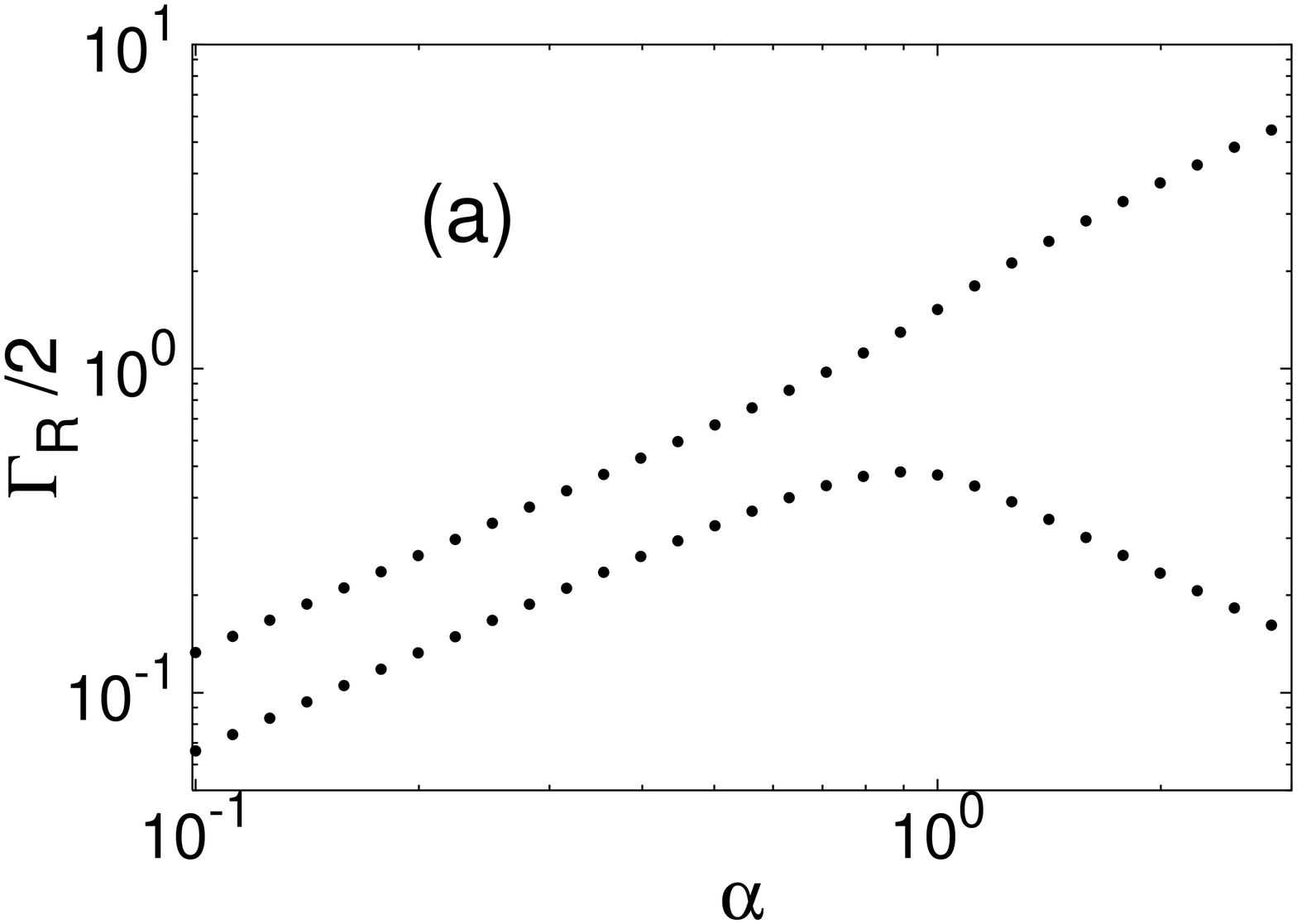}
\includegraphics[width=2.8cm,angle=-90]{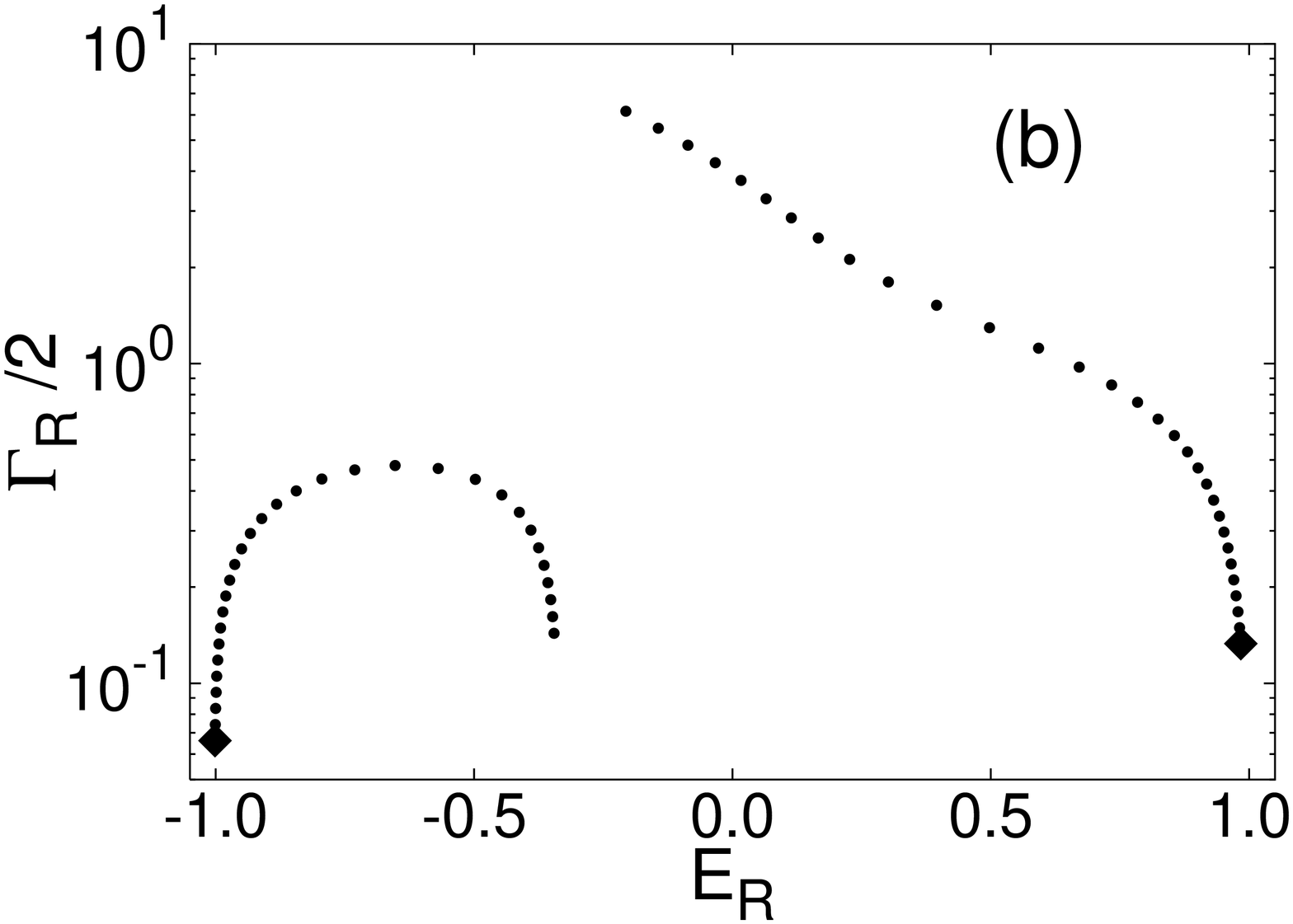}
\caption{Energy $E_R$ and width $\Gamma_R/2$ for two resonances in
the schematical model Eq.~(\ref{eq:twores}). {\bf (a)}
$\Gamma_R/2$ versus the coupling strength $\alpha$. {\bf (b)}
Trajectories of $E_R-i\Gamma_R/2$ as a function of $\alpha$.
The latter plot will be called eigenvalue picture in the
following.}
\label{fig:twores}
\end{figure}

Let us sketch the main ingredients of the trapping mechanism on
the basis of calculations for two neighboured resonances.  For
$N=2$, $K=1$, and $\alpha$ replaced by $\alpha e^{i\beta}$, the
Hamiltonian (\ref{eq:heff}) reads
\begin{equation}
H_2=
\left(
\begin{array}{cc}
 1 &  0 \\
 0 & -1
\end{array}
\right)
- 2i \alpha e^{i\beta}
\left(
\begin{array}{cc}
\cos ^2 \varphi & \cos \varphi \sin \varphi \\
\cos \varphi \sin \varphi & \sin ^2 \varphi
\end{array}
\right)
\label{eq:twores}
\end{equation}
where $\beta$ describes the ratio between the real and imaginary
parts of the coupling matrix, and $\varphi$ gives the relative
coupling strength of the two states. In Fig.~\ref{fig:twores}, the
widths $\Gamma_R/2$ and eigenvalues of the two resonances are shown
as a function of $\alpha$ for $\beta=\pi/18$ and $\varphi=\pi/5$.
For small $\alpha$, the widths of both states increase with
$\alpha$. Thereafter the two states attract each other in energy
and their widths bifurcate: The width of one state starts to
decrease with increasing $\alpha$ while the width of the other one
increases more strongly. At still larger $\alpha$, the broad
resonance gets shifted towards lower energies due to $\beta \neq
0$.

The goal of this paper is to present an experimental verification
of the effect of resonance trapping in a microwave cavity by
considering the local interactions between individual resonances.
The coupling of the discrete states of the cavity to an attached
waveguide makes the system open.
\begin{figure}
\includegraphics[width=8cm,angle=-180]{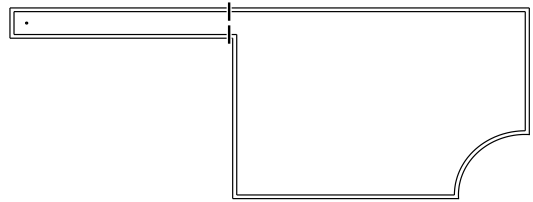}
\caption{Layout of the cavity (in scale): Quartered Sinai billiard
(285$\times$200 mm, radius 70 mm) with an attached waveguide
(220$\times$23.2 mm). The opening between waveguide and billiard
can be varied in steps of 0.1 mm.} \label{fig:cavity}
\end{figure}

The measuring technique is described elsewhere \cite{ste95}. Here
we note only that for flat cavities the electromagnetic spectrum
is equivalent to the quantum mechanical spectrum of the
corresponding system, as long as one does not surpass the
frequency $\nu_{\rm max}=c/2h$, where $h$ is the height of the
cavity. The quantum mechanical energy $E$ corresponds to the
square of the wavenumber $k$. For the measurement we used a
quartered Sinai billiard with an attached waveguide (see
Fig.~\ref{fig:cavity}).
The actual form of the cavity is of little importance since resonance trapping
is expected to take place in chaotic as well as in regular billiards
\cite{pepirose}.
The frequency range of the waveguide is
8.2 to 12.5 GHz. This corresponds to energies between $E=2.95 \,
{\rm cm}^{-2}$ and $6.85 \, {\rm cm}^{-2}$. Between the two
thresholds at $E=1.83 \,{\rm cm}^{-2}$ and $7.33 \, {\rm cm}^{-2}$
only one mode can propagate through the waveguide. The microwaves
are coupled into the system through an antenna at the end of the
waveguide. They enter the billiard through a slit, the opening of
which can be varied. The coupling matrix elements $V$, Eq.~(\ref{eq:heff}),
are related to the width of the opening. The exact expression for this
relation is not known for our situation with a limited number of resonances.
We performed 102 measurements for different
openings $d$ from $3$ to $23.2$ mm in steps of 0.2 mm. For all
openings we measured the relative amplitude and relative phase of
the reflected wave as a function of $E$ and expressed the values
through the complex reflection coefficient $R(E)$. Due to the
finite length of the waveguide with reflecting end there are also
broad channel resonances in the measured $R(E)$. This, together
with the fact that a part of the incoming flux gets absorbed in
the walls of the resonator, complicates the data analysis somewhat.

\begin{figure}
\noindent
\includegraphics[width=4cm,angle=-90]{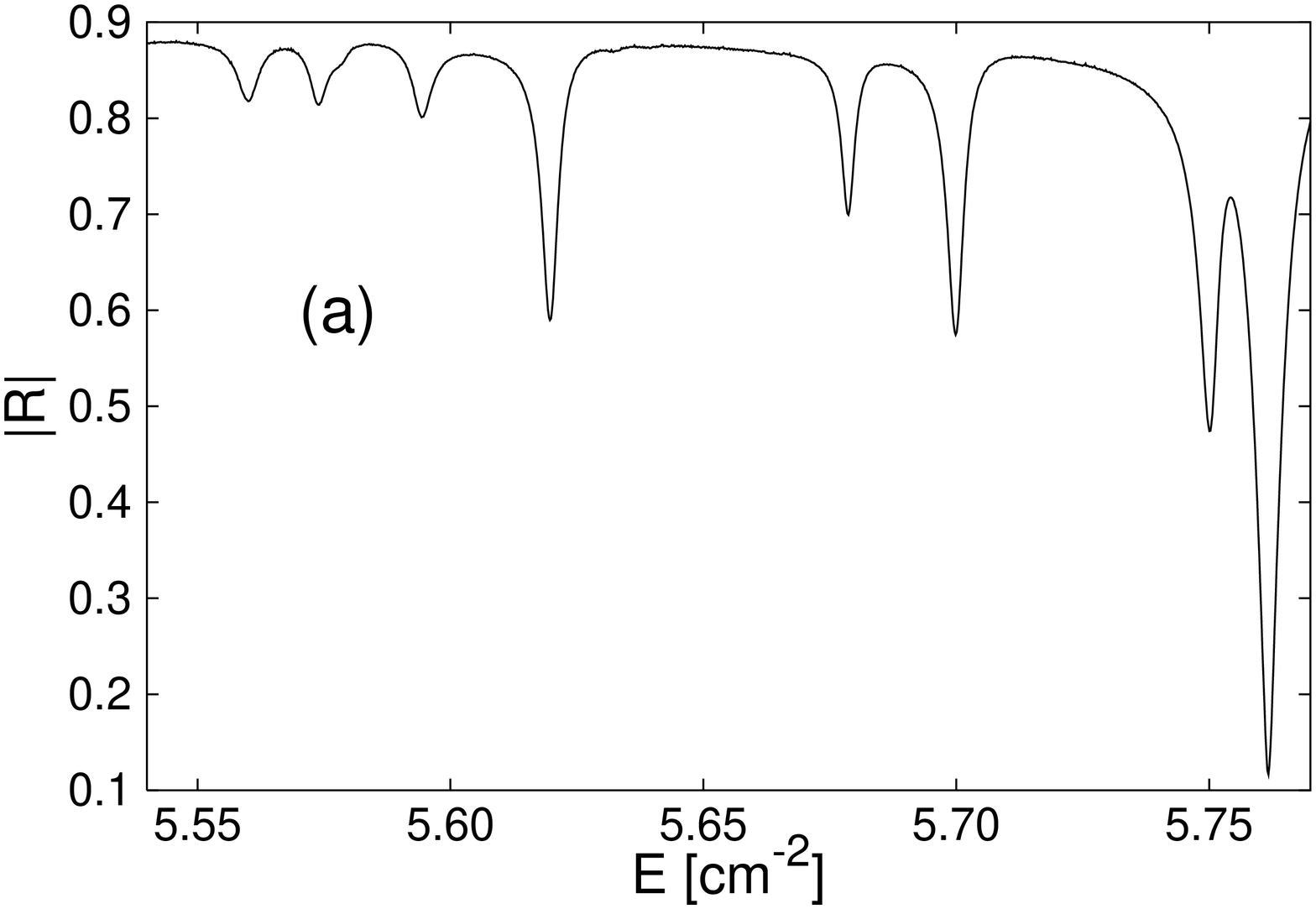}
\noindent
\includegraphics[width=4cm,angle=-90]{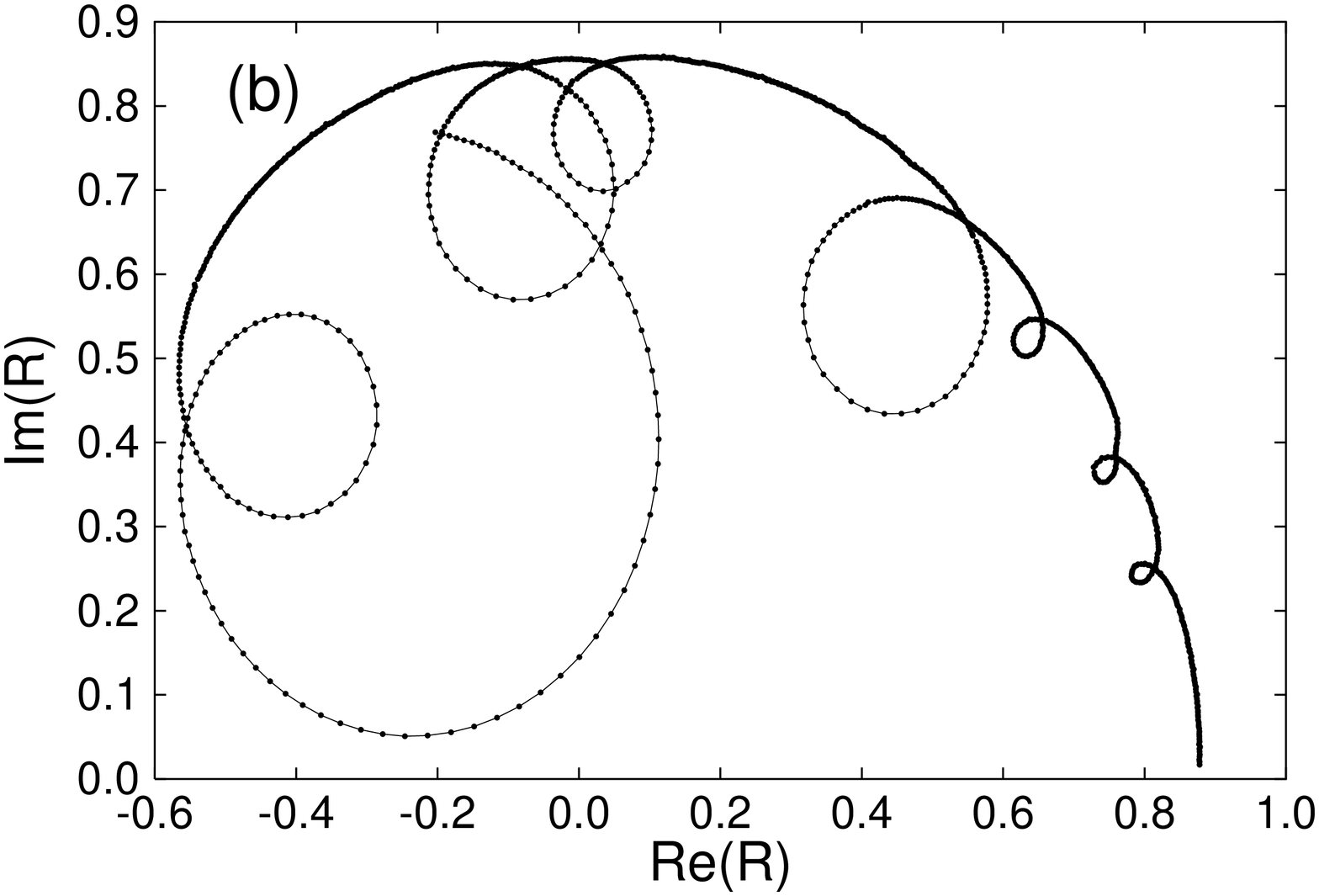}
\noindent
\includegraphics[width=4cm,angle=-90]{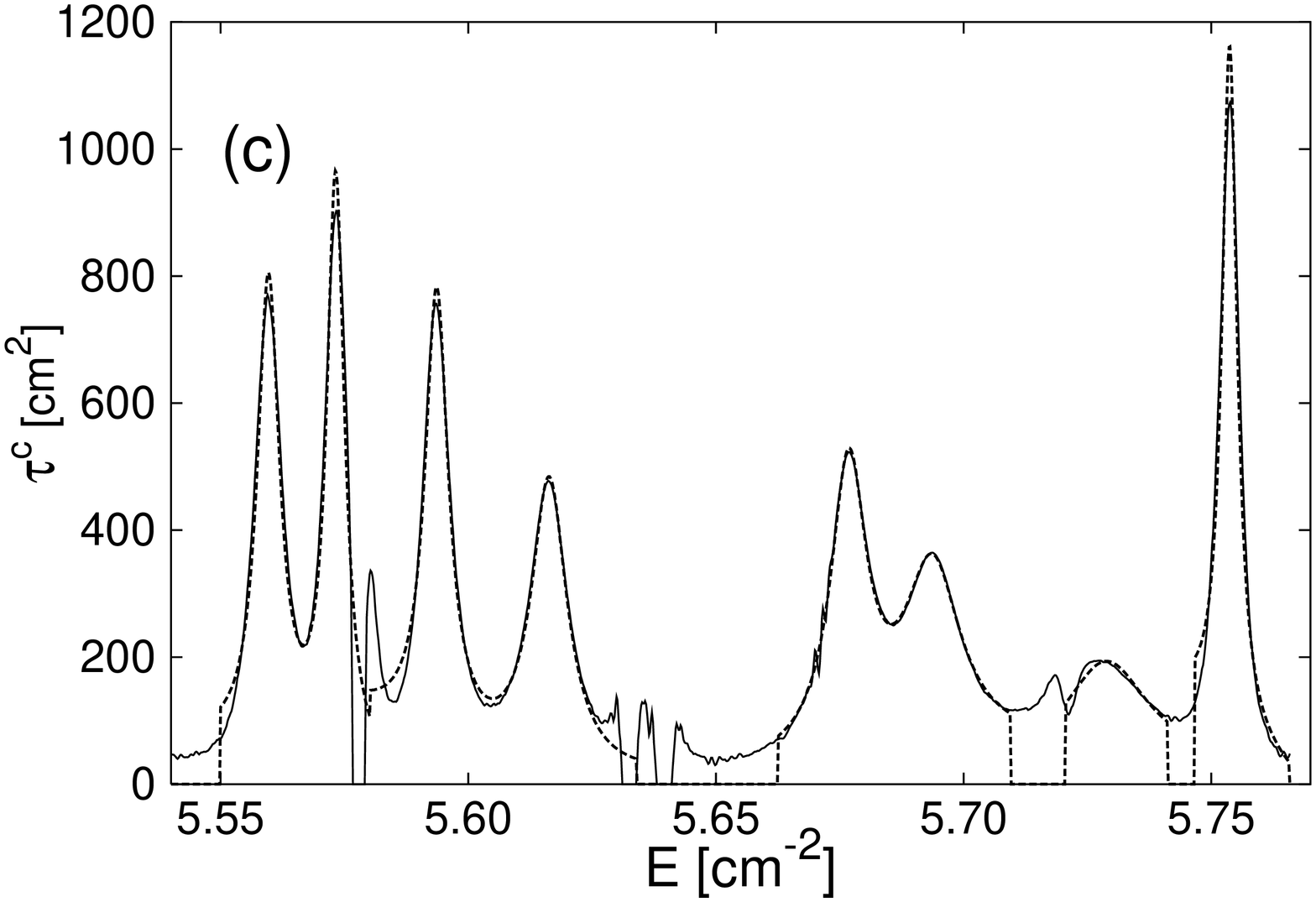}
\caption{{\bf (a)} Measured $|R|$ as a function of $E=k^2$. {\bf
(b)} Complex R for $5.54 \, {\rm cm}^{-2}\leq E \leq 5.77 \, {\rm
cm}^{-2}$ {\bf (c)} $\tau^c$ obtained directly from the
measurement (full line) and the fitted $\tau^c$ (dashed line).
In all three plots, the opening of the slit is $d=14$ mm.}
\label{fig:rr}
\end{figure}

\begin{figure}
\noindent
\includegraphics[width=5cm,angle=-90]{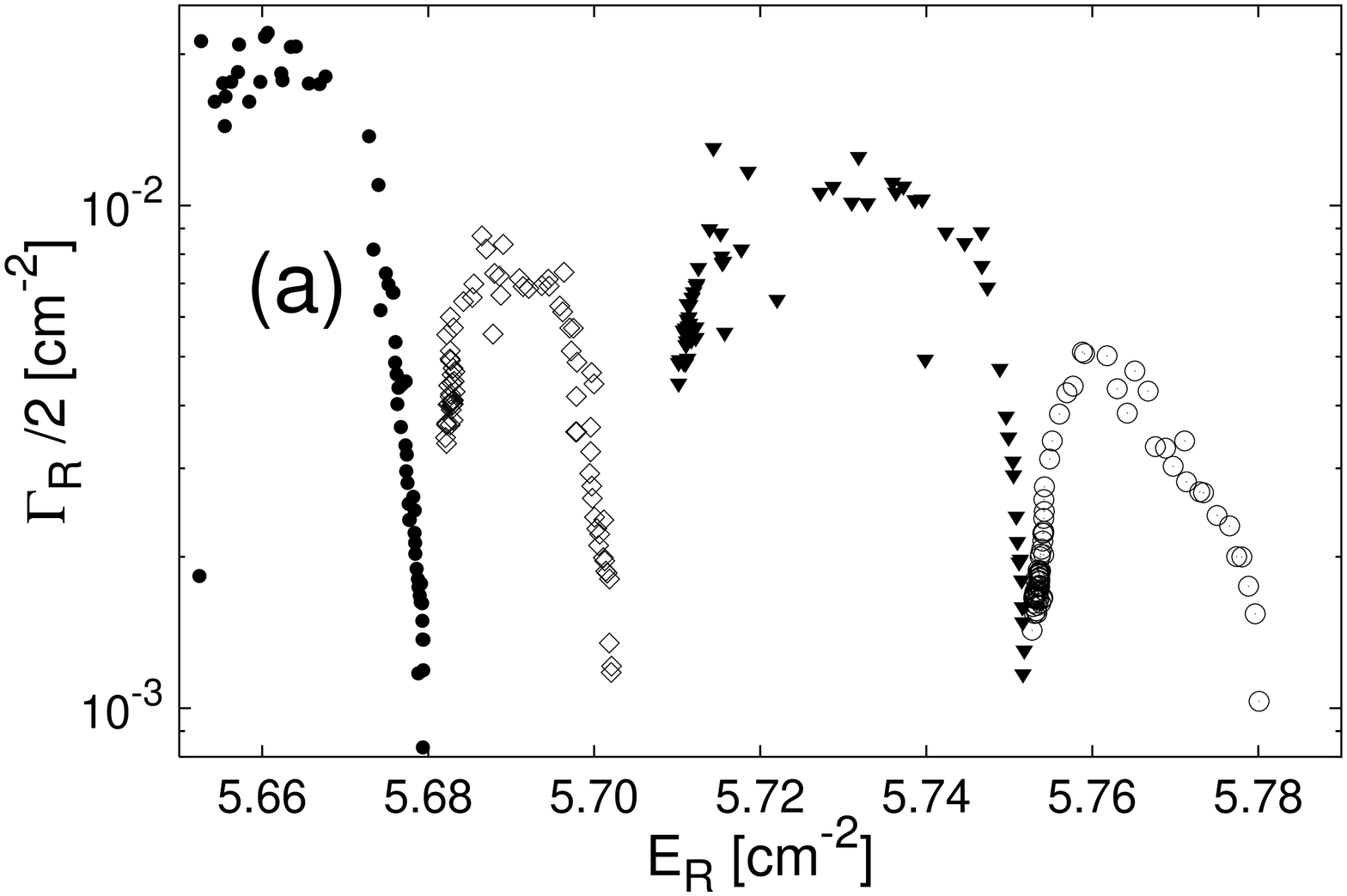}
\includegraphics[width=5cm,angle=-90]{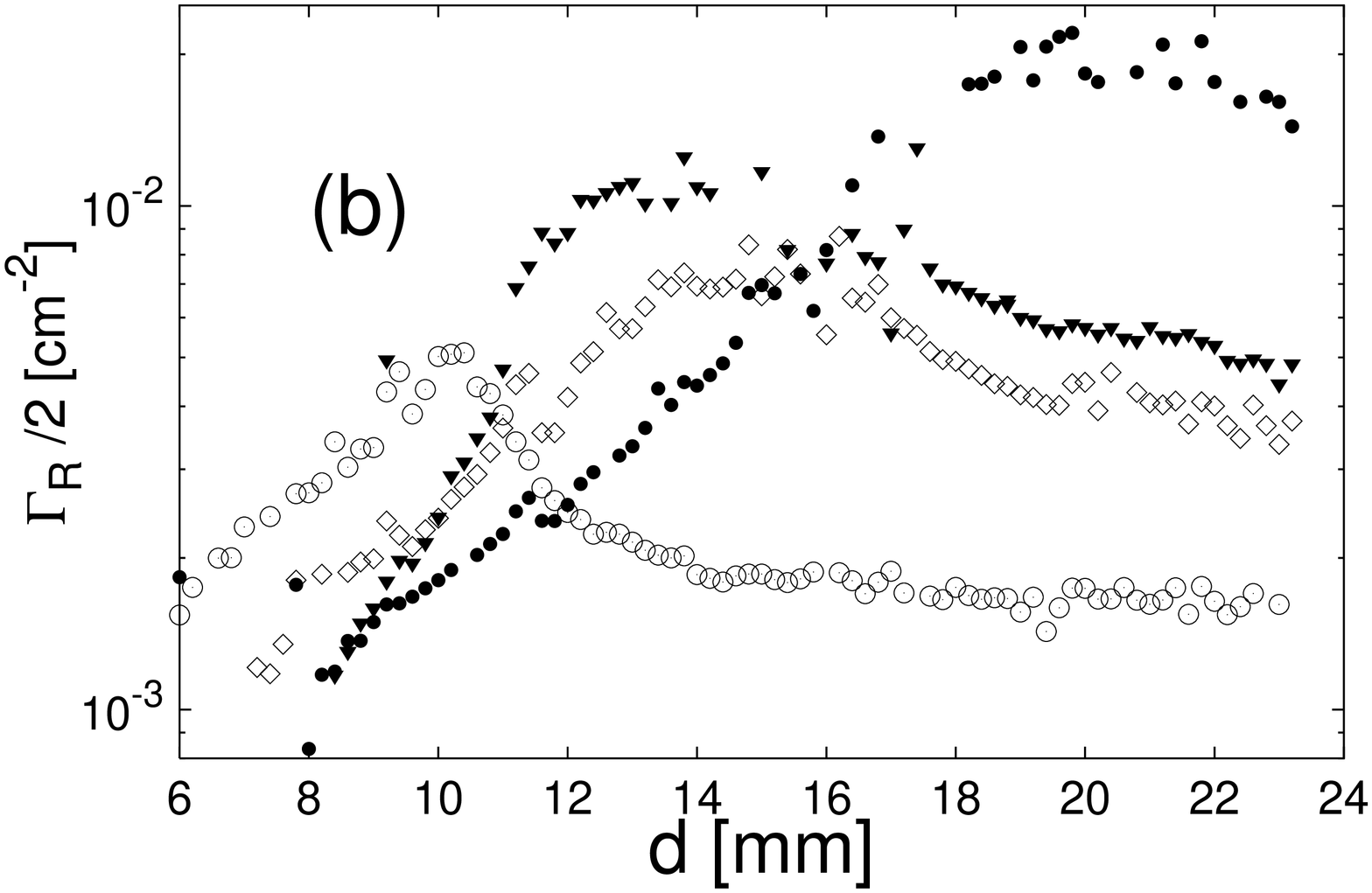}
\caption{Eigenvalue picture for four resonances {\bf (a)} and the
corresponding $\Gamma_R/2$ versus $d$ {\bf (b)}. The different
resonances are marked by different symbols. The widths of the resonances first
increase as a function of the opening $d$ of the slit but finally they
decrease again. This demonstrates the effect of resonance trapping.}
\label{fig:evpmagn}
\end{figure}

In Fig.~\ref{fig:rr}.a, a part of the measured spectrum $|R|$ for
$d=14$ mm is shown. Due to the wall absorption we have $|R|\leq 1$
and the resonances show up as dips in $|R|$. Around $E=5.75 \,
{\rm cm}^{-2}$ a pair of closely lying resonances can be seen. It
will be evident later (see Fig.~\ref{fig:evpmagn}) that resonance
trapping takes place between these two resonances.

The method used in this letter to obtain the energy positions
$E_R$ and widths $\Gamma_R$ of the resonances is as follows: In
Fig. \ref{fig:rr}.b we plot the real and imaginary part of $R(E)$
for $d=14$ mm in the region $5.54 \, {\rm cm}^{-2}\leq E \leq 5.77
\, {\rm cm}^{-2}$. This plot is known as the Argand diagram
\cite{argand}. In the figure, the narrow resonances show up as small circles
superimposed on larger circles caused by the broader structures. In order to
analyse the Argand diagram, we propose a centered time-delay
analysis (CTDA). From the
\noindent
measured points in a small region around an energy $E$, we
define a local circle segment. We calculate the center $C(E)$ of
this segment and define the angle $\theta^c(E)$ as the angle of
the complex number $R(E)-C(E)$. The point $C$ moves with $E$ in
such a way that $\theta^c(E)$ increases by $2\pi$ in the
neighbourhood of each resonance, even if the corresponding circle
does not go around the origin. We define the corresponding "time
delay" as $\tau^c(E)=d\theta^c(E)/dE$. For overlapping resonances,
$\tau^c$ is just the sum of the contributions $\tau_w^R$ from the
individual resonances. In the approximation of Breit-Wigner shaped
resonances we have,
\begin{equation}
\tau_w^R(E)=\Gamma_R/\left((E-E_R)^2+\Gamma_R^2/4\right) \; .
\label{eq:tauwr}
\end{equation}
In Fig.~\ref{fig:rr}.c, the $\tau^c$ obtained directly from the measurement
are shown for $d=14$ mm (full line). Further, we show the  $\tau^c$ (dashed
line) obtained from a fit to $\sum_R\tau_w^R(E)$,
Eq.~(\ref{eq:tauwr}). The measured data is split into smaller
energy intervals by searching for sufficiently small minima in
$\tau^c$. The fitted curve agrees well with the measured values.

The usual fit of the complex $R(E)$ to a sum of Lorentzians was
not possible because of the broad structures in $R(E)$ caused by
the channel resonances. By means of the CTDA, however, any broad
structures are automatically removed. It was also possible to do
an automatic evaluation of the measured data, which  was
mandatory in view of the large amount of data. To test the CTDA
method we performed an analysis of spectra created theoretically
from $H^{\rm eff}$, Eq.~(\ref{eq:heff}), with the wall absorption
simulated by adding a constant to the diagonal of the imaginary
part of $H^{\rm eff}$. The comparison between the theoretical
values and those extracted from the $R(E)$ showed a good
agreement. A more detailed study of the CTDA method together with
a comparison to possible other methods, e.g. the filter
diagonalization method \cite{mandel}, will be published elsewhere.

The $E_R$ and $\Gamma_R/2$ for four resonances obtained from the measured data
with the CTDA method are presented in Fig. \ref{fig:evpmagn}.
The motion of the $E_R$ and $\Gamma_R/2$ with
$d$ is clearly observable. At least three states start to
decrease in width with increasing $d$, i.\,e. there are some
evident cases of resonance trapping. Note that for each opening
$d$ of the slit the data have been fitted independently. The
smoothness of the curves is thus a measure for the reliability by
which the $E_R$ and $\Gamma_R$ are extracted from the measured data.
Fig. \ref{fig:evpmagn} even shows hierarchical trapping: one state
first traps its closest neighbour only to be trapped at stronger
coupling by another state.

We followed the motion of all the 127 resonance poles
$E_R-i/2\,\Gamma_R$ in the complex plane as a function of increasing $d$.
(Due to the small number of resonances, no comparison to statistical
predictions is made.) The widths of 47 resonances  decrease  with $d$
while those of 11 resonances stop to increase (within the accuracy
of the CTDA). These two groups can unambigously be identified as
trapped resonances, i.\,e. at least $46\%$ of the resonances get
trapped. For 16 of the resonances increasing further in width, it
was possible to identify other resonances getting trapped by them.
By this they aquired an extra contribution to their widths. In
total, we have found that at least $58\%$ of the resonances are
strongly affected by trapping. These resonances are in the whole
energy interval considered.

The estimation $58\%$ of the resonances getting affected by
resonance trapping is the lowest limit. According to theoretical
studies of the  diagonal elements of the effective Hamilton operator
Eq.~(\ref{eq:heff}) (i.e. without taking resonance trapping into account) and
studies for isolated resonances \cite{pepirose}, one would
expect a more or less uniform increase in width of all resonances.
The experiment shows, however, that the widths of the remaining
$42\%$ increase non-uniformly in $d$. This indicates that some of
these resonances gain (or loose) width at the cost (or in favour)
of other ones. Thus practically all of the resonances are affected
by the mechanism of resonance trapping.
\vspace{0.2cm}

In conclusion, we have demonstrated experimentally that resonance
trapping takes place in a microwave cavity coupled to a waveguide.
This is, to the best of our knowledge, the first unambigous
experimental verification of the effect. This  demonstration
was possible by tracing the motion of the resonance poles as a
function of the opening of the slit starting close to the real
axis and following them into the region of overlapping resonances.
Our experimental proof of resonance trapping
does not depend on any model assumptions.

Theoretical studies have shown that the phenomenon of resonance trapping
does neither depend on the number of resonances nor on the special shape of
the microwave cavity. The effect appears also in various open many-body
quantum systems. In any case, the decoupling of some states from the
decay channels  takes place. However, the experimental verification of
resonance trapping is important not only for an understanding of the
properties of open quantum systems with overlapping resonances.
More important is, maybe, the necessity of a
good knowledge of the detailed properties of open quantum systems
for the design of mesoscopic systems. By tuning the coupling
strength to the decay channels, the properties of the system can be
controlled. Further experimental and theoretical studies of this
interesting topic, including the conductance, have to be performed.
\vspace{0.1cm}

{\bf Acknowledgment:} The participation of Y.-H.~Kim,
R.~Sch\"afer, and H.~Schanze in performing the experiment is
gratefully acknowledged. E.~P. thanks the SFB 383 'Unordnung in
Festk\"orpern auf mesoskopischen Skalen' for taking the costs of a
three-month stay in Marburg. P.~Thomas and F.~Jahnke are thanked
for stimulating discussions.

\end{document}